\documentstyle[12pt,epsf]{article}
         \makeatletter
         \def\thefigure{\@arabic\c@figure}\def\fps@figure{tbp}
         \def\ftype@figure{1}\def\ext@figure{lof}
         \def\fnum@figure{\protect\footnotesize Fig.\ \thefigure}
         \def\thetable{\@arabic\c@table}
         \def\fps@table{tbp}\def\ftype@table{2}\def\ext@table{lot}
         \def\fnum@table{\protect\footnotesize Table \thetable}
         \makeatother
\begin{document}
\vspace*{0.3in}
\begin{center}

{\Large\bf
Collective Flow from the Intranuclear Cascade Model
}\\
\bigskip
\bigskip
David E. Kahana$^{1}$, Declan Keane$^{1}$, Yang Pang$^{2}$,
Tom Schlagel$^{2,3}$ and Shan Wang$^{1}$
\\
\vspace*{0.1in}
{$^1$\em 
Center for Nuclear Research, Physics Department\\
Kent State University, Kent OH 44240
}\\
\vspace*{0.1in}
{$^2$\em
Department of Physics, Brookhaven National Laboratory\\
Building 510A, Upton, NY 11973
}\\
\vspace*{0.1in}
{$^3$\em
State University of New York at Stony Brook\\
Stony Brook, NY 11794
}\\
\bigskip
\medskip
\end{center}
{\footnotesize
\centerline{ABSTRACT}
\begin{quotation}
\smallskip
\vspace{-0.10in}
The phenomenon of collective flow in relativistic heavy ion collisions is
studied using the hadronic cascade model ARC. Direct comparison is made to
data gathered at the Bevalac, for Au+Au at $p=1-2$
GeV/c. In contrast to the standard lore about the cascade model, collective
flow is well described quantitatively without the need for explicit mean field
terms to simulate the nuclear equation of state. Pion collective flow is
in the opposite direction to nucleon flow as is that of anti-nucleons
and other produced particles. Pion and nucleon flow are predicted at AGS
energies also, where, in light of the higher baryon densities achieved, we
speculate that equation of state effects may be observable.
\end{quotation}}
\bigskip
\bigskip
\bigskip

Collective flow$^{\ref{kitchen_sink}}$ in relativistic heavy ion collisions
has long been a subject of interest, since it was felt the phenomenon might
carry information about the nuclear matter equation of state$^{\ref{nuceos}}$.
In this letter, we will be concerned only with the so-called `Sideward Flow',
which is related to an event-by-event azimuthal asymmetry in the distribution
of final state particles.

Theories which have been used to describe the heavy ion collisions fall into
two broad classes: those based on macroscopic thermodynamical/hydrodynamical
considerations$^{\ref{hyd.mods.}}$, and those which attempt a more
microscopic description of the ion--ion collision, for instance by carrying
out successive collisions (cascading) of the elementary (hadronic)
constituents. Among the microscopic models one can distinguish pure cascade
models$^{\ref{Cugnon},\ref{ARC}}$, which include only the elementary $2$- or
in principle $n$-body collisions of the constituents. These models take as
their main input the experimentally measured cross-sections (and angular
distributions) for hadron-hadron $\sigma(hh\rightarrow X)$ in free space, and
then carry out the ion--ion collision by Monte-Carlo methods. The model,
{\bf ARC} = {\bf A} {\bf R}elativistic {\bf C}ascade$^{\ref{ARC}}$, which we
will use to discuss flow, is such a pure hadronic cascade model.

Additionally, there are microscopic models$^{\ref{VUU},\ref{RQMD}}$ which
include mean field, collective, or in-medium effects in some fashion, as well
as treating the elementary hadron--hadron collisions.
Both classes of microscopic model presumably simulate semi-classical
(relativistic) kinetic theory of the hadron gas in some limit, and these
models would then be equivalent to solving transport equations, either
Boltzmann-Vlasov, or Boltzmann-Vlasov-Uehling-Uehlenbeck (BUU), depending on
whether or not a mean field is included from the outset. In the BUU case, the
mean field enters through the gradient of a potential energy, which may
be calculated, e.g., from a phenomenological equation of state for nuclear
matter. 

Our treatment of flow using ARC will neglect mean fields entirely, and this
is based on the assumption that the mean field $U$ satisfies $U\ll T$ where
$T$ is the typical kinetic energy involved in a hadron--hadron collision
within the cascade. For the initial nucleon--nucleon collisions in Au+Au at
$p=1.7$ GeV/c this is a reasonable assumption, but of course the kinetic
energy available in successive collisions cascades down as the ion+ion
collision takes place. So, the mean field may not be negligible in late
or very soft collisions or for co-moving spectators in the projectile and
target. However, a spectator cut is applied to the data and to ARC so as to
reject particles having small kinetic energy in the projectile or target
frames. We expect therefore that mean field contributions will not be
dominant.

Indeed, we shall see that ARC, using un-modified free space cross-sections
and no mean field is, at least by direct calculation, adequate to the task of
describing sideward flow in Au+Au at Bevalac energies. Given the prior
extensive success of ARC$^{\ref{ARC2}}$ in predicting and describing
inclusive data for heavy ion collisions at AGS energies, our strong
theoretical prejudice would then be that mean fields and/or phenomenological
equations of state need not be included over this range of energies
(1-15 GeV/c). In-medium effects if and when they do arise ought then, in our
point of view, to be included by modifying the elementary interactions.
Nevertheless, the high baryon densities apparently achieved during massive
ion collisions at AGS energies$^{\ref{ARC}}$, may still manifest themselves
through traditional equation of state effects.

Our specific concern will be with the proton-like sideward flow measured at
the Bevalac in Au+Au collisions at lab momentum $p=0.96-1.9$ GeV/c. Such data
have already been measured using the Plastic Ball
spectrometer$^{\ref{plasticball}}$, and new experiments with better immunity
to detector distortions were recently carried out using the EOS time
projection chamber; flow results from the EOS collaboration are expected to
be available later this year$^{\ref{eoscoll}}$. We anticipate the EOS data
by considering forward rapidity data only in our comparisons, as is reasonable
given the downstream location of the EOS detector. By proton-like, we mean to
say that protons contained in identified outgoing nuclear fragments are
counted towards the flow, together with outgoing free protons. We consider
proton-like flow, because ARC does not as yet dynamically include the
production of nuclear fragments larger than single nucleons. Coalescence
calculations for deuterons and tritons$^{\ref{HIPAGS-deuterons}}$ from ARC
have been carried out at AGS energies, and are in good agreement with data.
There is in principle no obstacle to carrying these calculations to lower
energy.

We calculate sideward flow {\it \`a la} Danielewicz and
Odyniec$^{\ref{flow_def}}$, by first defining a reaction plane for the
ion+ion collision, neglecting pions, using the beam direction and a vector
defined as a weighted average of outgoing transverse momenta:
\def\Q{{\bf Q}}
\def\p{{\bf p}}
\begin{equation}
\Q = \sum_{i} w(y_i) \p_T^i, 
\end{equation}
with $w(y)$ selected as in Ref($\ref{flow_def}$). The essential point is that
$w$ is an odd function of $y_{cm}$. The sideward flow curve is then defined
by projecting proton-like momenta into the reaction plane, and averaging:
\begin{equation}
\langle P_x(y) \rangle = \frac{\sum_i (\p_i \cdot \Q_i)/|\Q_i|}{N(y)}
\end{equation}
\noindent Here $N(y)$ is the number of protons detected in a bin of width
$dy$ around $y$. Kinematic cuts are placed on the ARC calculation. These are:
\newcounter{cut}
\begin{list}%
	{(\arabic{cut})}{\usecounter{cut}\setlength{\rightmargin}{\leftmargin}}
	\item Spectator Cut: $p_{proj} > 0.25$ MeV.
	\item Forward Rapidity Cut: $y_{cm} > 0$
\end{list}
\noindent The cuts are chosen with the yet to be published EOS data in mind.
The target in the EOS detector is located upstream of the main
tracking chamber, which optimises acceptance for $y_{cm}>0$ but compromises it
for $y_{cm}<0$. This combination of cuts ensures that our flow calculation is
made in a region where detector distortions should be minimal$^{\ref{eoscoll}}$.
The charge multiplicity $M$ is then defined as the number of protons
surviving the cuts.

ARC was designed, of course, for much higher energy heavy-ion collisions
at AGS, so it was not certain from the outset how well lower energy Bevalac
data would be described. However, we felt it was of interest to compare flow
from the Plastic Ball and the forthcoming EOS data, especially the recent Au+Au
runs at maximum Bevalac energy, with an unmodified version of ARC. We do not
include comparisons of inclusive spectra, because such data have not been
published for the Plastic Ball, and are not yet published in a journal in the
case of EOS. We simply comment that the agreement with mid-rapidity spectra
in Ref(${\ref{spectra}}$) seems very good considering that the present beam
energy is a factor of ten below that for which ARC was designed. Differences
between theoretical and experimental spectra can be expected to show up at
forward rapidities due to a simplified treatment of fermi motion which does
not completely respect conservation of energy. This is of course a negligible
effect at the AGS energies.

Our flow comparison is based on the slope of the $\langle P_x\rangle$ curve
near mid-rapidity. This observable is minimally distorted by the Plastic Ball
acceptance, as judged by simulations. The measured slope for Plastic Ball
events with a multiplicity selection comparable to that placed on ARC (the
upper $50$\% of events in the multiplicity spectrum are accepted) is shown as
a dotted line in Fig(1). The Plastic Ball slope is corrected for dispersion
in the estimated reaction plane; for the calculation of ARC
$\langle P_x\rangle$ we use the ideal reaction plane. 

It is commonly asserted that cascade models produce little or no collective
flow$^{\ref{early_cascade}}$. ARC, like any cascade model, generates
$\langle P_x \rangle$ through successive collisions, and the details of the
treatment of these individual collisions may increase or decrease the amount
of flow produced. Two salient features of ARC (as well as of other cascade
models) with respect to flow are:
\newcounter{points}
\begin{list}%
	{(\arabic{points})}{\usecounter{points}
			\setlength{\rightmargin}{\leftmargin}}
	\item ARC treats two body collisions in the center of momentum frame,
	      where the impact parameter at closest approach and the initial
	      momentum together span a plane: the two body reaction plane.
	      For those collisions also having two bodies in the final state
	      cascades often generate the azimuthal angle (relative to the
	      reaction plane) of the outgoing momentum randomly.
	\item After final state momenta are chosen, there still exist
	      distinct possibilities for the virtual spatial path of the
	      particles through the collision, corresponding semi-classically
              to repulsive and attractive orbits. Cascade models typically
              allow either type of orbit, choosing at random between the two.
\end{list}

Point (1) is not consistent, classically, with the conservation of angular
momentum, which requires initial and final momenta to lie in the same
plane. As for point (2), it is well known that the nucleon--nucleon
interaction is repulsive at high energies, where one expects to see only the
hard core. Moreover, at the energy of first collisions here, ($T_{cm}=450$
MeV), and in most ensuing collisions, there should not be a large attractive
component to the scattering. Therefore two possible sources of flow suggest
themselves: preserving the $2$-body reaction plane, and eliminating attractive
orbits. 

\begin{figure}
  \vbox{\hbox to\hsize{\hfil

		\epsfxsize=4.5truein
		\epsffile[83 129 537 737]{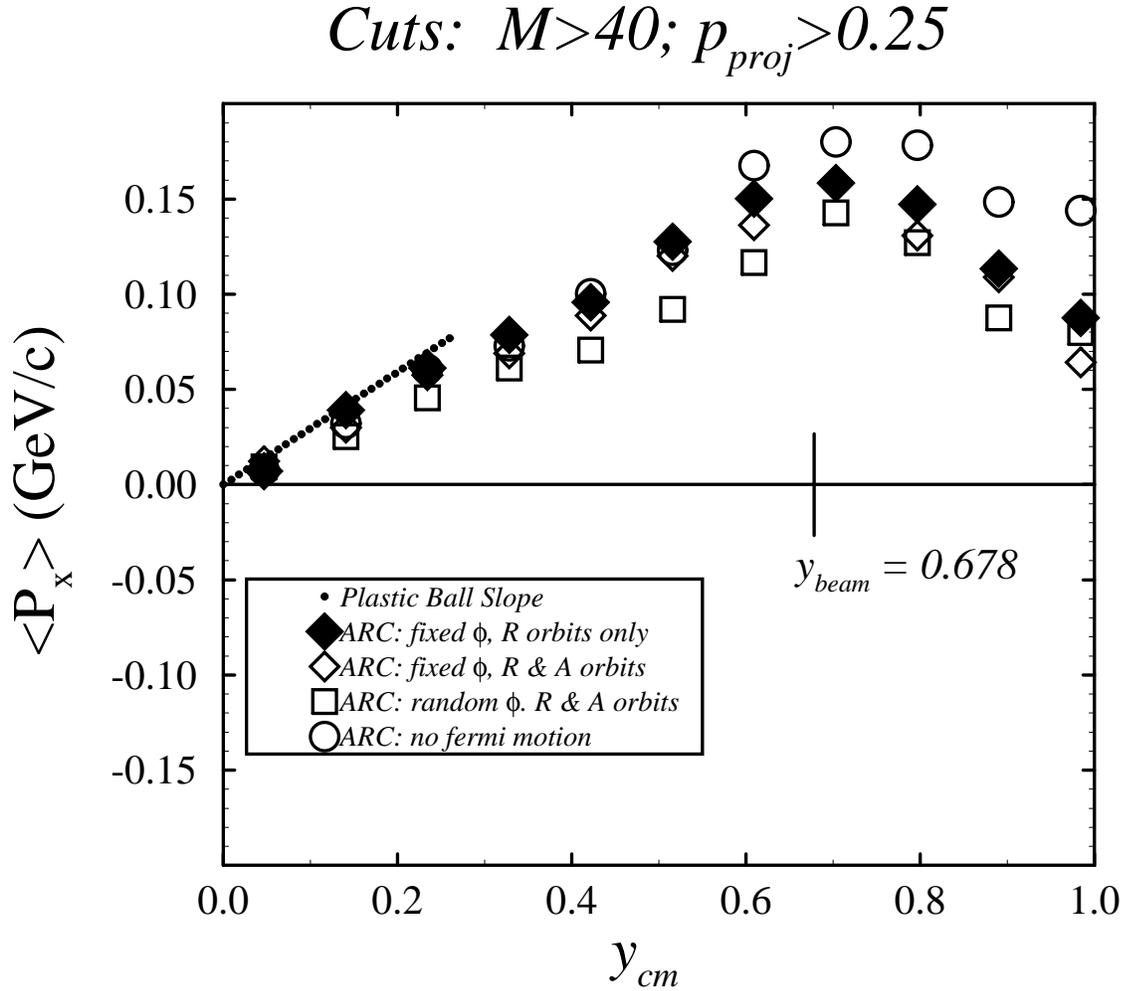}
		\hfil}
		}

  	\caption[]{\footnotesize
	FLOW: PLASTIC BALL vs. ARC (Au+Au at $1.7$ GeV/c). Dotted line
	indicates Plastic Ball slope at mid-rapidity, corrected for
	dispersion in the estimated reaction plane. The line is not
	extended to forward rapidity since the Plastic Ball acceptance
	filter has a significant effect there. ARC $\langle P_x \rangle$
	is calculated in the ideal reaction plane.

}
\label{Fig:flow_eos_vs_arc}
\end{figure}

Results from ARC, for a beam momentum of $1.7$ GeV/c, are plotted in
Fig(\ref{Fig:flow_eos_vs_arc}). One can see that accounting for the final
state plane adds $\sim 20$\% to the average flow, while eliminating
attractive orbits produces an additional $\sim 10$\%, or so. We have also made
calculations for beam momenta of $0.96$ and $1.9$ GeV/c in order to explore
the dependence of these results on collision energy. The agreement between
theory and experiment remains equally good at these higher and lower energies.

Some further theoretical discussion is perhaps in order here, since
the cascade model mixes classical propagation between collisions, with
empirical scattering cross-sections. If justification can be found in quantum
mechanics for the cascade model, it is in part from the eikonal approximation
to potential scattering, valid when $U\ll T$, and for small angle scattering.
This is a generalisation of the Born approximation, which assumes that
scattering particles move through the potential along a straight line path or
classical orbit (WKB); thus one obtains an impact-parameter-dependent phase
shift for the outgoing wave. A generalisation of the eikonal approximation to
the $n$-body system with pairwise interactions would retain the notion of an
impact parameter for the $2$-body collisions.  Hence a $2$-body reaction
plane exists, and should be preserved for two-body final states.  Quantum
mechanically the angular momentum must still be conserved, and so a $2$-body
plane is reasonably well-defined, as long as the scattering is not in an
$s$-wave. We expect a spread in the angle of the plane: $\Delta\theta \sim
(2l + 1)^{-1}$, and a simple estimate can be made of the number of partial
waves present, for the beam momentum and a typical cross-section: $l_{max} =
kb \sim 5$, ($b=(\sigma/\pi)^{1/2}$), for the first collision. In this way one
could allow for quantum mechanical smearing of the orientation of the two
body plane according to the angular momentum present, but a first
approximation is just to preserve the plane exactly. We expect that not much
of the scattering is $s$-wave at $p_L\sim 1.7$ GeV/c, even for ensuing
cascading collisions.

One further issue of concern is whether fermi motion, which as we pointed out
above could be handled more correctly, is in fact somehow responsible for the
flow exhibited by the cascade. This question can be dispensed with
immediately, by examining Fig(1).  One sees that, if anything, the flow
increases when fermi motion is turned off.

We expect that some small adjustment of the theory may result when
consistent fermi motion, smearing of the $2$-body plane, and
impact-parameter-dependent choice of repulsive or attractive orbits are
introduced, but certainly there would not appear to be any need at this point
for large mean field terms, and collective flow at these energies can, it
seems, come from a pure cascade model.

\begin{figure}
  \vbox{\hbox to\hsize{\hfil

		\epsfxsize=4.5truein
		\epsffile[83 129 537 737]{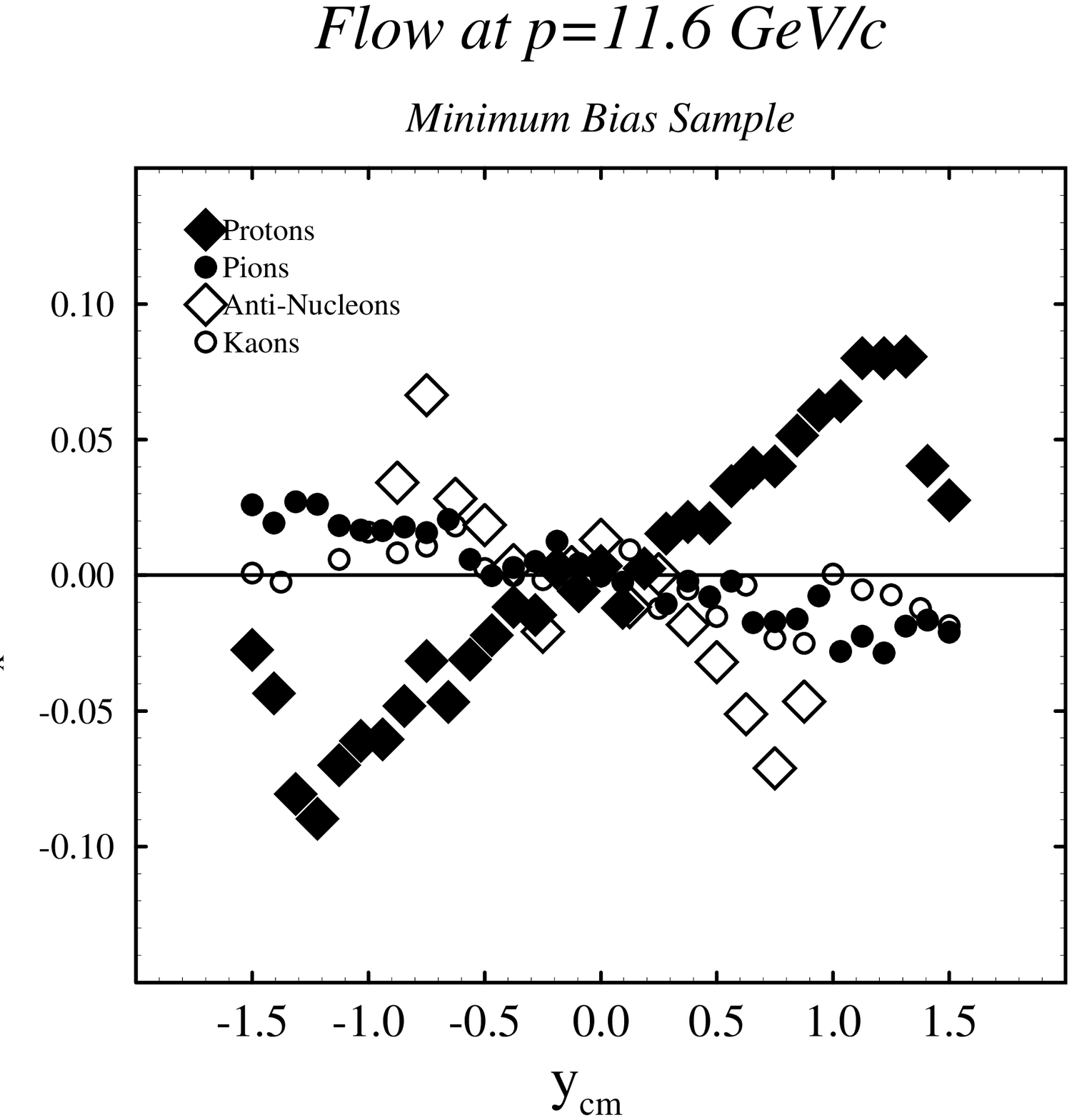}
		\hfil}
		}

  	\caption[]{\footnotesize
	FLOW at AGS ENERGY: Pion and nucleon flows are shown, as well as
	kaon and anti-nucleon flows. Produced particles all have flow
	momentum in the opposite direction to the nucleons.  	   			}
	\label{Fig:pion_flow}
\end{figure}

It is relevant to ask what the pions do in relation to the nucleons.  The
direction and size of flow in the pion sector might give a hint about the
underlying mechanism generating $\langle P_x \rangle$ in the nucleons.
Overall momentum conservation, involving both positive and negative rapidity,
does not constrain the direction, or size, of any pion flow. Pursuing this
question, we calculated flow at AGS energies, however without using specific
experimental cuts. For the pions we define a sideward flow, based on the
reaction plane already determined from the protons. A graph of the pion and
proton flow curves, obtained in Au+Au at $p=11.6$ GeV/c is shown in
Fig(\ref{Fig:pion_flow}). It is seen that the pion flow is in the opposite
direction to the proton flow at these energies, which suggests strongly that
pion production enhances the nucleon flow.
We observe that the flow momentum for all produced particles is in the opposite
direction to that for nucleons; flow curves for anti-nucleons, kaons, and pions
in Fig(\ref{Fig:pion_flow}) make this evident. Though more study is
needed, it seems that the magnitude of anti-nucleon flow is similar to that
for protons, while kaon and pion flows are somewhat less.

In peripheral collisions, that is to say, in collisions having appreciable
sideward flow, particle production will take place initially in the dense
central region where target and projectile nuclei overlap. Anti-nucleons
trying to emerge from the interaction region in the direction of nucleon flow
will encounter more target or projectile nucleons than those emerging in the
opposite direction. Given the large annihilation cross-section, it is natural
that anti-nucleons should {\it anti-flow}. Of course, anti-nucleons are a
small perturbation to the overall dynamics, but a similar mechanism could
function in the case of kaons and pions, where absorption on nucleons is
still a strong effect. However, many low energy pions, from $\Delta$ decay,
follow the nucleon flow.

Nucleon flow may also arise from nucleons scattering inelastically at the
edge of and away from the dense central region of the collision, giving an
average transverse momentum to the produced pions, in the opposite
direction. At the kinetic energy ($T_{cm}\sim 3.5$ GeV) of first $pp$
collisions, the cross-section is more inelastic than at lower energies. So,
production plays an important role at the higher energies. Even at the beam
momentum of $1.7$ GeV/c ($T_{cm}=450$ MeV), considered for the Bevalac Au+Au
data, production is relevant. However, a new dynamical region may be entered
below the threshhold for pion production. In $pp$ the inelastic cross-section
has fallen essentially to zero when the center of mass kinetic energy is
below $200$ MeV. Then, the mechanism for flow production will involve only
elastic processes and mean fields may perhaps make significant contributions.

That a pure cascade model can produce sufficient collective flow is a
surprising and interesting result. The question might be asked: why did early
cascade models fail to produce enough flow? From the outset, it should be
stated that the literature is somewhat contradictory on this question.  Early
cascade model calculations can be found displaying up to 50\% of
experimental flow in various systems$^{\ref{early_cascade}}$. It is crucial to
know, when evaluating results, whether experimental cuts have been correctly
applied to the simulations, since numerical results for flow are quite
sensitive to these cuts. For instance, the magnitude of the flow calculated
here would be significantly larger for a spectator cut of $0.35$
GeV/c. Nevertheless, the general conclusion seems to have been arrived at,
that the pure cascade generates little or no collective flow.  We would argue
that this conclusion is incorrect at least at Bevalac energies, and above, and
that a correctly constructed cascade does in fact produce enough flow without
mean fields. As we have noted above, mean fields presumably still play some
role as the collision energy is reduced.

One may also ask: if the equation of state need not be included explicitly in
the modeling to generate flow, then is any role played by the equation of
state in producing flow? Certainly the cascade model, including as it does
classical, relativistic kinetic processes, produces at least the thermal
pressure and equation of state of an ideal relativistic gas. This thermal
pressure may be all that is needed for flow. At the very least the present
calculation suggests an insensitivity to off-shell aspects of the equation of
state.

At AGS energies, flow is predicted in both pions and nucleons, and it is a
not inconsiderable effect. The direction of sideward flow is opposite in
pions and nucleons. The success of ARC in describing flow at Bevalac
energies, and inclusive data at AGS energies emboldens us to use
the code to study flow at AGS energies. High energy flow predictions
hopefully will be testable when the EOS detector is moved to BNL and the E895
collaboration begins to take data. Since the densities achieved are higher
than at the Bevalac, we can still hope to see explicit equation of state
effects, and flow may be a good observable to study in this connection. One
may look for deviations from the pure cascade picture, in rarer, very high
multiplicity events.

The authors thank S.H.Kahana for useful discussions. This work has been
supported by DOE grants No. DE-FG02-89ER40531, DE-FG02-93ER40768,
DE-AC02-76CH00016, and NSF grant No. PHY91-13117. 
\bigskip
\bigskip
\vfill\eject
\noindent{\bf References}
\smallskip
\begin{enumerate}
\vspace{-10pt}
\item \label{kitchen_sink}
        H.\ \AA.\ Gustafsson {\it et al.}, Phys.\ Rev.\ Lett.\ {\bf 52},
	1590 (1984).
\vspace{-10pt}
\item \label{nuceos}
	For reviews, see H. St\"ocker and W. Greiner, Phys. Rep. {\bf 137},
	277(1986);\\
	G. F. Bertsch and S. Das Gupta, {\it ibid.} {\bf 160}, 189 (1988). 
\vspace{-10pt}
\item \label{hyd.mods.}
	H. St\"ocker, J. Maruhn, and W. Greiner, Z. Phys. {\bf A293}, 173
	(1979);	Phys. Rev. Lett. {\bf 44}, 725 (1980).\\
	J.\ Kapusta and D.\ Strottman, Phys. Lett. {\bf B106}, 33, (1981).
	277(1986);\\
	G. F. Chapline, {\it et al.}, Phys. Rev. {\bf D8}, 4302 (1973).\\
	W. Scheid, H. M\"uller, and W. Greiner, Phys. Rev. Lett. {\bf 32},
	741 (1974).\\
	C. Y. Wong and T. A. Welton, Phys. Lett. {\bf 49B}, 243 (1974).
\vspace{-10pt}
\item \label{Cugnon}
	J. Cugnon, Phys. Rev. {\bf C22}, 1885 (1980).\\ 
	R. Serber, Phys. Rev. {\bf 72}, 1114 (1947).\\
	Y. Yariv and Z. Fr\"ankel, Phys. Rev. {\bf C20}, 2227 (1979);
	{\it ibid.} {\bf 24}, 488 (1981).
\vspace{-10pt}
\item \label{ARC}
	Y. Pang, T. J. Schlagel, and S. H. Kahana, Nucl.\ Phys.\
	{\bf A544}, 435c (1992), {\it ARC -- A Relativistic Cascade}, 
	Proceedings of Quark Matter'91.
\vspace{-10pt}
\item \label{VUU}
	G. Bertsch, S. DasGupta, and H. Kruse, Phys. Rev. {\bf C29}, 673
	(1984);\\ 
	J. Aichelin and G. Bertsch, Phys. Rev.  {\bf C31}, 1730 (1985).\\
	H. Kruse, B. V. Jacak, and H. St\"ocker, Phys. Rev. Lett. {\bf 54},
	289 (1985);\\
	J. J. Molitoris and H. St\"ocker, Phys. Rev. {\bf C32}, 346 (1985).
\vspace{-10pt}
\item \label{RQMD}
	R. Mattiello, {\it et al.},
	Phys.\ Rev.\ Lett. {\bf 63}, 1459 (1989);\\
	H. Sorge,  {\it et al.}, 
	Phys.\ Rev.\ Lett. {\bf 68}, 286 (1992).\\
	J. Aichelin and H. St\"ocker, Phys. Lett. {\bf B176}, 14 (1986).
\vspace{-10pt}
\item \label{ARC2}
        Y. Pang, T. J. Schlagel, and S. H. Kahana, Phys. Rev. Lett.,
	{\bf 68}, 2743, (1992).\\
        T. J. Schlagel, S. H. Kahana, and Y. Pang, Phys. Rev. Lett., 
	{\bf 69}, 3290, (1992).
\vspace{-10pt}
\item \label{plasticball}
	K. G. R. Doss {\it et al.}, Phys. Rev. Lett. {\bf 57}, 302 (1986);\\
	K.-H. Kampert, Nucl. Part. Phys. {\bf 15}, 691 (1989);\\
	H.~H.~Gutbrod, A.~M.~Poskanzer, and H.~G.~Ritter,
	Rep. Prog. Phys. {\bf 52}, 1267 (1989).
\vspace{-20pt}
\item \label{eoscoll}
	M. Partlan {\it et al.}, LBL-, submitted for publication. 
\vspace{-10pt}
\item \label{HIPAGS-deuterons}
	A.\ J.\ Baltz, {\it et al.}, Proceedings HIPAGS '93, MIT,
	January 1993.
\vspace{-10pt}
\item \label{flow_def}
	P.\ Danielewicz and G.\ Odyniec, Phys. Lett. {\bf 129B}, 146 (1985).
\vspace{-10pt}
\item \label{spectra}
	G. Rai {\it et al.}, to appear in Proc. of {\it International
	Workshop on Dynamical Features of Nuclei and Finite Fermi Systems},
	(World Scientific, Singapore, in press).
\vspace{-10pt}
\item \label{early_cascade}
	D.\ Beavis {\it et al.}, Phys.\ Rev.\ {\bf C33}, 1113, (1986).\\
\vspace{-10pt}
\end{enumerate}

\end{document}